\documentclass[11pt,a4paper]{article}
\usepackage{graphicx}
\usepackage{multirow,booktabs}
\usepackage{array}
\usepackage{makecell}
\usepackage[font=small,labelfont=bf,textfont=it]{caption}
\usepackage{microtype}

\begin{document}
\begin{center}
\noindent\textbf{\Large On event-by-event pseudorapidity fluctuations in relativistic nuclear interactions}\\[4mm]
\textbf{M. Mohisin Khan$^{1*}$, Danish F. Meer$^1$, Tahir Hussain$^2$, N. Ahmad$^3$} \\[2mm]
 \textit{1. Department of Applied Physics, ZHCET, Aligarh Muslim University, Aligarh, India}\\
 \textit{2. Applied Sciences and Humanities Section, University Polytechnic, Aligarh Muslim University, Aligarh, India}\\
\textit{3. Department of Physics, Aligarh Muslim University, Aligarh, India}\\[2mm]

\textbf{$^{*}$mohsinkhan.ph@amu.ac.in}\\[2mm]

\textbf{Abstract}
\end{center}
Present study is an attempt to have a detailed look into event-by-event(e-by-e) pseudorapidity fluctuations of the relativistic charged particles produced in $^{28}$Si-nucleus interactions at incident momenta 4.5A and 14.5A GeV/c. The method used in the present study makes use of a kinematic variable which is derived in terms of the average pseudo-rapidity and the total number of particles produced in a single event. The multiplicity and pseudorapidity dependence of these fluctuations have also been studied. The results obtained for the experimental data are compared with HIJING simulation.\\


\noindent\textbf{Introduction}\\
Relativistic nuclear collisions are the most fascinating and important tools to produce matter under extreme conditions of temperature and density. The key point to study and understand the behaviour of this produced matter is the copious production of secondary particles in these collisions.The global observable such as multiplicity and pseudo-rapidity of produced particle play an important role in understanding the particle production process in the relativistic hadron-hadron, hadron-nucleus and nucleus-nucleus collisions. The current interest in such studies are mainly to understand the characteristics of quark-gluon plasma (QGP) and the scenario of phase transition from QGP to the normal hadronic phase. Fluctuations in the values of global observable have always been considered as one of the possible signature of QGP formation$^1$. Various nearly conclusive studies$^{2-5}$ regarding QGP formation and its signatures have been made using the data on three experimental energy regime from the SPS, RHIC and LHC.\\As the matter produced in high energy heavy-ion collisions is a short live state and the hadronization after the collisions is very fast one has to rely on the observations made on the characteristics of the produced particles. The study of the particles coming out of the interaction region may provide important information regarding the underlying dynamics of collision process and the multi-particle production. Fluctuations in general and the event-by-event fluctuations of observable in particular are envisaged to give vital information about the phase transition$^{5-9}$. The process of thermalization along-with the statistical behaviour of the produced particles can be understood by studying the fluctuation in particle multiplicity and momentum distribution$^{10-15}$. Reference 6 stressed that the charge fluctuations may be an evidence of QGP formation. Many such studies about the fluctuations has been carried out. However, study of critical point of QGP phase transition can be well study using the concept of e-by-e fluctuations because in this case the fluctuations are predicted to be large enough $^{14-17}$. A study of each event produced in relativistic nuclear collision may reveal new physical phenomena occurring in some rare events for which conditions might have been created in these collisions. Nuclear collisions at high energies produce a large number of particles, the analysis of single event with large multiplicity can shed light on some different physics than the study of averages over a large events. Predictions have been made about the occurrence of critical density fluctuations in the vicinity of the phase transition and its manifestation as e e-by-e fluctuation of different physical observable $^{18}$. The e-by-e analysis may offer a possibility of observing a phase transition directly if the particle emitting source is hydro-chemical composition. The NA49 collaboration$^{10}$ observed the fluctuations of transverse momentum and koans to pins ratio in central Pb-Pb collisions at 158A GeV. A. Bialas and V. Koch
$^{9}$ and Belkacem et al$^{19}$ reported the moments of e-by-e fluctuations are very nearly related  to the correlation function. The ALICE collaboration$^{20}$ has measured the e-by-e fluctuation in mean transverse momentum in p-p and Pb-Pb collisions at LHC. A number of papers are available in literature on e-by-e fluctuation analysis of different observable but a very few papers are there on e-by-e pseudo-rapidity fluctuations. The first such study was carried out by the KLM collaboration for 158A GeV Pb-AgBr interactions$^{21}$, Recently S. Bhattacharya et al.$^{18}$ and Gopa Bhoumic  et al.$^{22}$ has carried out the e-by-e pseudo-rapidity fluctuations analyses on various emulsions data having different projectiles and targets at 4.1A GeV, 4.5A GeV, 60A GeV and 200A GeV.\\
In the present study we have carried out e-by-e fluctuations analysis for the data at 4.5A and 14.5A GeV $^{28}$Si-AgBr interactions for the experimental and HIJING simulated data.\\
Following sections of this paper are devoted to the  details of the data, analysis methods, results and discussion and the observations made on the basis of obtained results.\\\\
\noindent\textbf{Experimental details of the data}\\
The present analysis has been carried out on the experimental and simulated data. For the experimental data, two random samples consisting 555 events of 14.5A GeV/c $^{28}$Si-nucleus interactions and 530 events of 4.5A GeV/c $^{28}$Si-nucleus interactions with $N_s \ge $10 have been used where N$_s$ represents the number of charged particles produced in an event with relative velocity ($\beta = v/c > 0.7$). The emission angles of the relativistic charged particles were measured and their pseudo rapidities ($\eta = -ln (tan(\theta /2)$) are determined. All other details about the data may be found elsewhere$^{23,24}$. Furthermore, for comparing the experimental results with the corresponding values obtained for the events generated by Monte Carlo code HIJING-1.33$^{25}$ event generator, a similar sample of events was simulated.\\

\noindent\textbf{Method of analysis}\\
M. Gazdzicki and S. Mrowczynski$^7$ proposed an excellent method to measure the fluctuations of any global kinematic variable. It is worth-mentioning that the second moment of distributions of global kinematic variables (multiplicity, rapidity, transverse momentum etc.) for individual events or all the events taken together may shed light on the extent of thermalization and randomization feature of high energy nuclear collisions. The basic idea used in this method is the fact that the correlated production of particles in each elementary interaction leads to large e-by-e fluctuations and these fluctuations  in high energy nuclear collisions are believed to originate due to trivial variation in impact parameter of the interaction. It may also be aroused due to some statistical reasons or due to some dynamical reason of the underlying processes prevailing at the instant of the collisions. The method proposed here$^7$ automatically filters the trivial contributions and provides a way to determine the remaining part contributing to the fluctuations. For this, a variable, $\Phi$ which is believed to be a measure of fluctuation is defined whose nonzero values points towards the correlation and fluctuation and a vanishing value of $\Phi$ points towards the independent particle emission(random emission) from a single source. The detailed procedure of calculating this variable is described below.\\
As the global kinematic variable used in the present analysis to study the e-by-e fluctuation is the pseudo-rapidity,$\eta$, of the emitted particles we first define a single particle variable z in terms of $\eta$ as:

\begin{equation}
z= \eta-\bar{\eta},
\end{equation}
where $\bar{\eta}$ represents the mean value of single particle inclusive pseudo-rapidity distribution that can be expressed as
\begin{equation}
\bar{\eta}=\frac{1}{N_{total}}\sum_{m=1}^{N_{evt}}\sum_{i=1}^{N_m} \eta_m ,
\end{equation}
where N$_m$ is the multiplicity of m$^{th}$ event. The second summation over i in the above equation  runs over all the particle produced in the m$^{th}$ event and the first summation is performed over all the events N$_{evt}$ in the sample. N$_{total}$ in the denominator is the total number of particles produced in all the events. Further, a multi-particle analogue of z, Z$_k$, is defined as
\begin{equation}
Z_k=\sum_{i=1}^{N_k}\eta_i-\bar{\eta}.
\end{equation}
Finally, the measure of fluctuation parameter, $\Phi$ is defined as
\begin{equation}
\Phi=\sqrt{\frac{<Z^2>}{<N_{total}>}}-\sqrt{\bar{z^2}},
\end{equation}
where the $<Z^2>$ and $<N_{total}>$  represents the event averaged of the variables therein and the $\sqrt{\bar{z^2}}$ is the square root values of the second moment of inclusive z distribution. As stated$^7$, $\Phi$ vanishes when there is no correlation among the produced particles and its non vanishing values are a measure of correlations and fluctuation present in the system. This method has been extensively used with success to analyze many experimental data$^{18,21}$ and to verify various aspects theoretically$^{26,27}$. In the present analysis we have attempted to study e-by-e fluctuation in 4.5A and 14.5A GeV/c $^{28}$Si-nucleus interactions. As the present analysis is meant to study the e-by-e $\eta$ fluctuations, the variable z is defined as \\~\\

\noindent\textbf{Results and discussion}\\
First of all the values of $\Phi$s are calculated for different groups of events selected on the basis of the average multiplicity of the relativistic charged particles, $<N_s>$. This calculation has been made for both the experimental and simulated data. These values along with statistical errors are tabulated in Table 1. The the different groups are selected in such a way to ensure that the average multiplicity of the group is greater than the average multiplicity of the sample of the data.

\begin{table*}[!htb]\centering
  \caption{Calculated values of $\Phi$s for different multiplicity classes for the experimental and HIJING simulated data.}
\renewcommand{\arraystretch}{1.2}
\setlength{\tabcolsep}{4.0pt}
\begin{tabular}{@{}c c c c c c@{}}
  \toprule[0.8pt]

  Interactions &  \makecell {Multiplicity\\selection} & \multicolumn{2}{c}{Experimental} & \multicolumn{2}{c}{HIJING} \\
  \cmidrule[0.3pt](r){3-4} \cmidrule[0.3pt](l){5-6}
  & & $<N_s>$ &  $\Phi$ & $<N_s>$  & $\Phi$ \\ \midrule[0.5pt]

  \multirow{5}{*}{\makecell{4.5A GeV/c\\ $^{28}$Si-nucleus}} & N$_s$ $\ge$ 10 &    15.33 & 5.402 $\pm$ 0.073 & 17.85 & 4.491 $\pm$ 0.088\\
  & N$_s$ $\ge$ 20  & 31.54 & 5.010 $\pm$ 0.071  &  33.55 & 3.923 $\pm$ 0.080\\
  & N$_s$ $\ge$ 30  & 42.14 & 4.410 $\pm$ 0.082  &  39.74 & 3.108 $\pm$ 0.082\\
  & N$_s$ $\ge$ 40  & 52.76 & 4.106 $\pm$ 0.088  &  51.23 & 2.213 $\pm$ 0.089\\
  & N$_s$ $\ge$ 50  & 64.75 & 3.710 $\pm$ 0.069  &  63.25 & 1.984 $\pm$ 0.094\\
 \midrule[0.5pt]
  \multirow{5}{*}{\makecell{14.5A GeV/c\\ $^{28}$Si-nucleus}} & N$_s$ $\ge$ 10 &  24.98 & 5.281 $\pm$ 0.074 & 19.71 & 3.874 $\pm$ 0.089\\
  & N$_s$ $\ge$ 20  & 33.99 & 5.010 $\pm$ 0.077  &  36.25 & 3.093 $\pm$ 0.101\\
  & N$_s$ $\ge$ 30  & 47.99 & 4.740 $\pm$ 0.088  &  43.55 & 2.823 $\pm$ 0.113\\
  & N$_s$ $\ge$ 40  & 51.86 & 3.901 $\pm$ 0.097  &  50.55 & 2.123 $\pm$ 0.134\\
  & N$_s$ $\ge$ 50  & 64.66 & 2.558 $\pm$ 1.066  &  61.58 & 1.674 $\pm$ 0.149\\  \bottomrule[0.8pt]
  \end{tabular}
  \label{tab:1}
\end{table*}

Tabulated  above table are the values of $\Phi$ and the averaged multiplicity of relativistic charged particles, $<N_s>$ for various multiplicity classes for the experimental and simulated events. It is observed from the table that the values of $\Phi$ are non-zero for all the multiplicity classes considered in the present study and this observation is same for both the experimental and simulated data at the two incident energies for $^{28}$Si-nucleus interactions. These non zero values of $\Phi$  supports the occurrence of dynamical fluctuations in the $\eta$-variable and the presence of correlation during particle production process in high energy nucleus-nucleus collisions. It is also observed from the table that the $\Phi$, which is considered to be the strength of correlation and fluctuation, shows a decreasing trend with increasing $<N_s>$. This dependence of $\Phi$ on $<N_s>$ is shown in Fig.1. The errors shown in the figure 1 are the statistical one. It is clear from Fig1. that the e-by-e $\eta$ fluctuation tends to decrease  with increasing mean multiplicity of the produced relativistic charged particles. This may be due the fact that the contribution to particle production would have been taken place by several independent sources. These contribution might be masking the correlated production. One can argue that there are identical sources which are producing low multiplicity events which is resulting in the low $\Phi$ values. It means when source fluctuation tends to vanish, pseudo-rapidity fluctuation increases. This observed trend of variation of $\Phi$ with the average multiplicity at high energies is in agreement with observations made by other studies in high energy regime$^{18,20,21}$.

\begin{figure}
  \includegraphics[width=\textwidth,keepaspectratio=true]{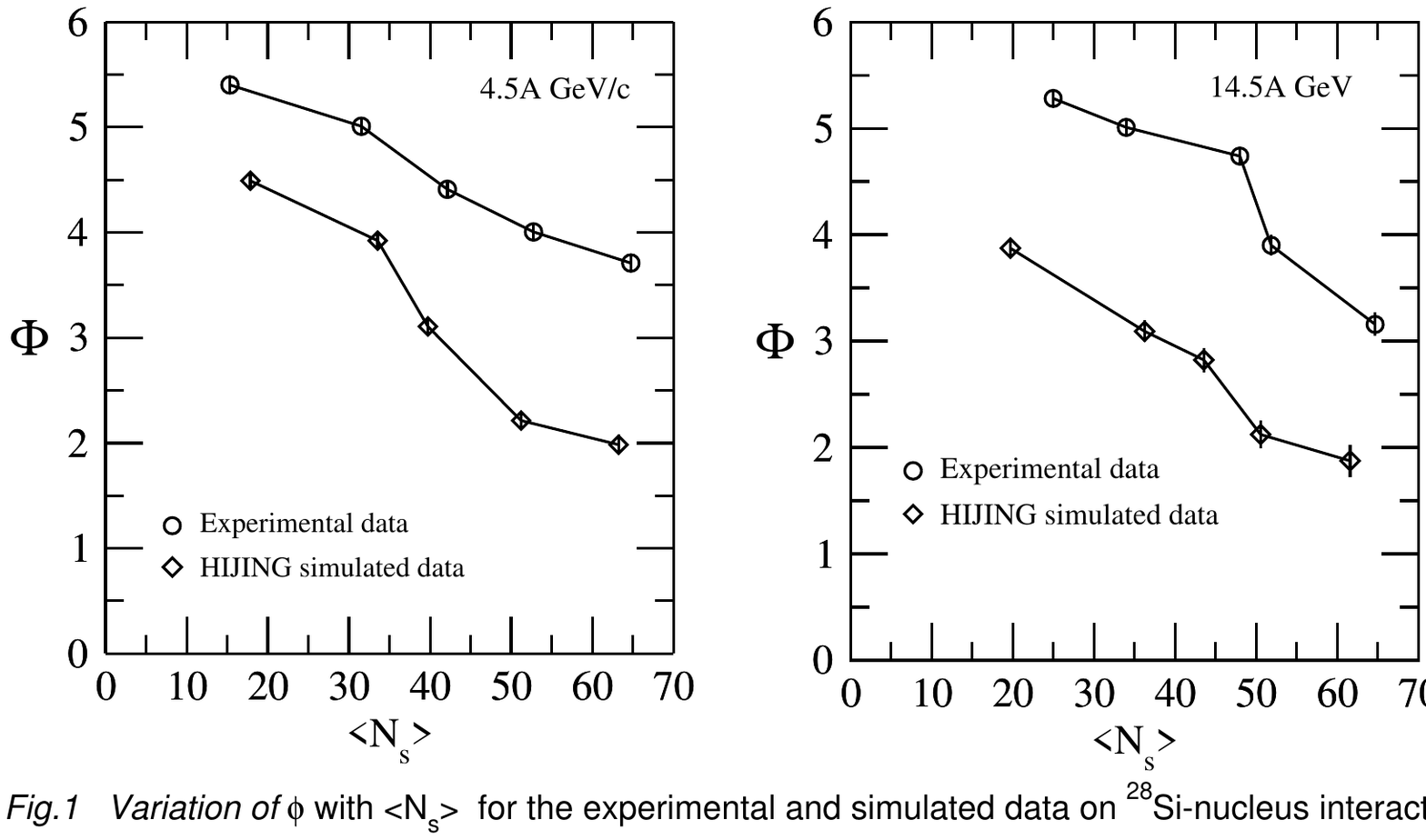}
\end{figure}

To compare the experimental results with HIJING simulation, sample of events are generated with statistics approximately 10 times that of the experimental statistics. It is clear from Fig.1 that the $\Phi$ values obtained for HIJING data are lower than the its values for the experimental data at both the energies but the trend of variation of $\Phi$ with $<N_s>$ is almost same for both experimental and simulated data.\\

Another interesting aspect of the event-by-event pseudo-rapidity fluctuation or the fluctuation of any global observable describing high energy nuclear collision data is to see its dependence on phase space region, which in this study is the pseudo-rapidity space itself. For this, the values of $\Phi$s are determined for various pseudo-rapidity intervals, $\Delta\eta = \eta_2 - \eta_1$, where $\eta_1$ and $\eta_2$ are the lower and upper limit of a chosen $\eta$-window. In the present study the chosen $\Delta\eta$ are 0.5,1.0,1.5,2.0,3.0,4.0,5.0,6.0 for both the data sets. The calculated values of $\Phi$s corresponding to these pseudo-rapidity regions are listed in Table 2.

\begin{table*}[!htb]\centering
  \caption{Calculated values of $\Phi$s in different pseudo-rapidity windows for the experimental and HIJING simulated data.}
\renewcommand{\arraystretch}{1.2}
\setlength{\tabcolsep}{25pt}
\begin{tabular}[\textwidth]{@{}c c c c@{}}
  \toprule[0.8pt]
Interactions & $\Delta\eta$ & \multicolumn{2}{c}{$\Phi$} \\
\cmidrule[0.3pt]{3-4}
& & Experimental & HIJING  \\ \midrule[0.5pt]

\multirow{9}{*}{\makecell{4.5A GeV/c\\ $^{28}$Si-nucleus}} & 0.5 & 0.199 $\pm$ 0.009 & 0.116 $\pm$ 0.008\\
& 1.0 & 0.499 $\pm$ 0.018 & 0.362 $\pm$ 0.023 \\
& 1.5 & 1.159 $\pm$ 0.028 & 0.905 $\pm$ 0.082 \\
& 2.0 & 2.179 $\pm$ 0.041 & 1.937 $\pm$ 0.088 \\
& 2.5 & 2.890 $\pm$ 0.098 & 2.987 $\pm$ 0.092 \\
& 3.0 & 3.972 $\pm$ 0.101 & 3.257 $\pm$ 0.098 \\
& 4.0 & 4.452 $\pm$ 0.161 & 4.096 $\pm$ 0.117 \\
& 5.0 & 4.622 $\pm$ 0.201 & 4.362 $\pm$ 0.188 \\
& 6.0 & 5.027 $\pm$ 0.211 & 4.674 $\pm$ 0.198 \\
\midrule[0.5pt]
\multirow{9}{*}{\makecell{14.5A GeV/c\\ $^{28}$Si-nucleus}} & 0.5 & 0.194 $\pm$ 0.006 & 0.102 $\pm$ 0.021 \\
& 1.0 & 0.387 $\pm$ 0.009 & 0.341 $\pm$ 0.029 \\
& 1.5 & 1.097 $\pm$ 0.021 & 0.891 $\pm$ 0.038 \\
& 2.0 & 2.063 $\pm$ 0.082 & 1.912 $\pm$ 0.043 \\
& 2.5 & 2.732 $\pm$ 0.111 & 2.889 $\pm$ 0.055 \\
& 3.0 & 3.817 $\pm$ 0.128 & 3.172 $\pm$ 0.076 \\
& 4.0 & 4.158 $\pm$ 0.141 & 3.995 $\pm$ 0.102 \\
& 5.0 & 4.489 $\pm$ 0.188 & 4.355 $\pm$ 0.111 \\
& 6.0 & 4.811 $\pm$ 0.214 & 4.788 $\pm$ 0.175 \\ \bottomrule[0.8pt]
\end{tabular}
  \label{tab:2}
\end{table*}

\noindent It is observed from Table 2 that as we widened the pseudo-rapidity space, we noticed larger e-by-e fluctuations. The values of $\Phi$s first increases with increasing $\Delta\eta$ and then tends to saturate for much larger $\Delta\eta$. This behaviour is observed at both the energies considered in this analysis. This might be due to the dominating long-range correlations as compared to short-range correlations as we explore a larger rapidity space. Based on phenomenological evidence, it has been argued that particle production in high energy hadron-hadron and nucleus-nucleus collisions have been carrying the signals of both the short and long range correlations. The average number of produced particles virtually depend on the size of the initiating cluster, this gives rise to the long range correlation, means the particles which are separated by relatively large $\eta$ shows some correlation. The values of $\Phi$s for HIJING simulated data are smaller as compared to its values for the experimental data but HIJING data too show the similar dependence of $\Phi$ on $\Delta\eta$. These observations are much more clearly depicted in Fig.2, where we plotted $\Phi$ against $\Delta\eta$ along-with statistical errors.\\

\begin{figure}
  \includegraphics[width=\textwidth,keepaspectratio]{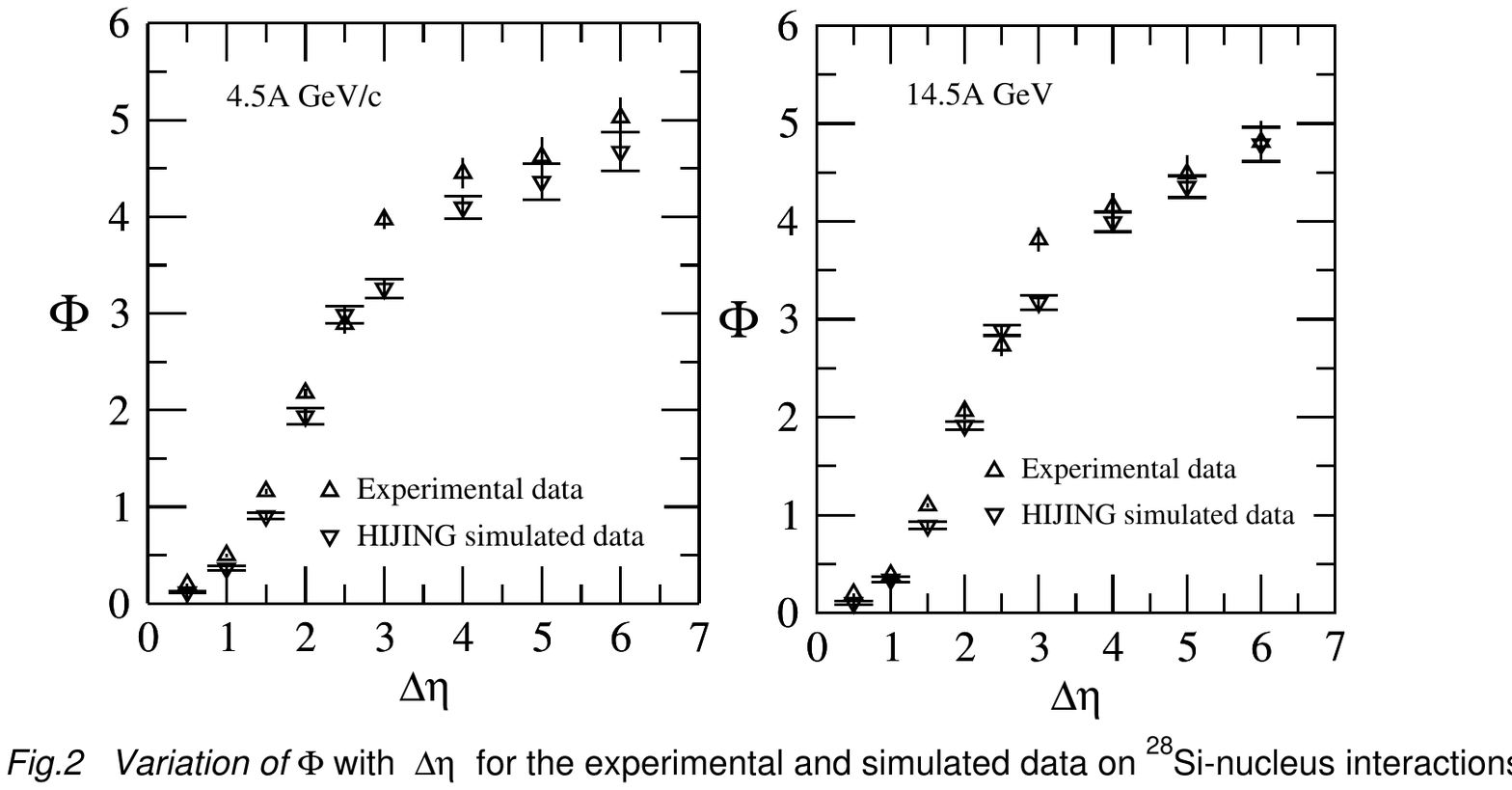}
\end{figure}

\noindent\textbf{Conclusions}\\
Event-by-event fluctuations of pseudo-rapidity of the relativistic charged particles produced in $^{28}$Si-nucleus interactions at 4.5A and 14.5A GeV/c have been studied in terms of the fluctuation and correlation quantifying parameter, $\Phi$. Analysis reveals the presence of e-by-e $\eta$-fluctuations and correlation amongst the produced particles in pseudo-rapidity space at both the incident momentum as the non vanishing values of $\Phi$s are obtained. It's observed that these fluctuations decreases with increasing mean multiplicities of the produced particles. This might be due to smearing out of the existing correlation as the more and more independent particle emitting sources added up. E-by-e fluctuations are also found to depend on the pseudo-rapidity windows and shows an increasing behaviour with increasing $\Delta\eta$. Results obtained for HIJING data exhibit a similar trend as compared to the experimental data at both the energies. Correlation and fluctuation studies remain to be excellent tools to explore the behaviour of the system produced in heavy-ion collisions at relativistic and ultra-relativistic energies.\\~\\

\noindent\textbf{Acknowledgment:}
Financial support from DST, Govt. of India is acknowledged with thanks.\\~\\

\noindent\textbf{References}
\begin{enumerate}
\item M. A. Stephanov, K. Rajagopal, E.V. Shuryak, Phys. Rev. Lett. \textbf{81} 4816(1998)
\item M. Luzum eta al., J. Phys. G. \textbf{41} 063102 (2004)
\item Y.Yoki et al., Nature \textbf{443}, 675 (2006)
\item S. Jeon et al., Phys. Rev. C \textbf{73}, 014905 (2006)
\item L. F. Babichev, A.N. Khmialeuski, \textit{Proceeding of 15$^{th}$ Int. Conf.-School}, September 20-23, 2010
\item M. Weber for the ALICE collaboration, J. Phys.: Conf. Series \textbf{389}, 012036 (2012)
\item M. Gazdzicki, S. Mrowczynski, Z. Phys. C \textbf{54}, 127 (1992)
\item E.V. Shuryak, Phys. Lett B \textbf{423}, 9 (1998)
\item A. Bialas, V. Koch, Phys. Lett. B \textbf{456}, 1(1999)
\item NA49 Collaboration (H. Appelshauser et al.), Phys. Lett. B \textbf{459}, 679 (1999)
\item G. Baym, H. Heisenberg, Phys. Lett. B \textbf{469}, 7 (1999)
\item G. Danilov, E. Shuryak, nucl-th/9908027
\item T. Anticic et al., Phys. Rev. C \textbf{70}, 034902 (2004)
\item T. K. Nayak, J. Phys. G. \textbf{32}, S187 (2006) arXiv:nucl-ex/060802.
\item H. Heiselberg, Phys. Rep. \textbf{351}, 161(2001)
\item M. Stephanov et al., Phys. Rev. Lett. \textbf{81}, 4816 (1998)
\item M. Stephanov et al., Phys. Rev. D \textbf{61}, 114028 (1999)
\item S. Bhattacharya et al., Phys. Lett. B \textbf{726}, 194 (2013)
\item M. Belkacem et al., arXiv:nucl-th/9903017v2, 22 April 1999
\item B. Abelev et al. Eur. Phys. J. C \textbf{74}, 3077 (2014)
\item KLM Collaboration (M.L. Cherry et al.), Acta Phys. Pol. B \textbf{29}, 2129(1998)
\item Gopa Bhoumic, Swarnapratim Bhattacharya et al., Euro.Phys.J. A\textbf{52} 196(2016)
\item Shafiq ahmad et al., J. Phys. Soc. Jpn., \textbf{75}, 064604 (2006)
\item Shakeel Ahmad et al., Acta Phys. Pol. B \textbf{35}, 809 (2004)
\item M. Gyulassy and X.N. Wang, Comp. Phys. Commun., G \textbf{25} 1895 (1999)
\item M. Gazdzicki et al., Eur. Phys. J. C\textbf{6}, 365 (1999)
\item M. Mrowczynski, Phys. Lett. B\textbf{439} 6 (1998)
\end{enumerate}

\end{document}